\documentclass{article}

\usepackage{spconf,amsmath,graphicx}
\usepackage{cite}
\usepackage{booktabs} % For professional looking tables: \toprule, \midrule, \bottomrule
\usepackage{multirow} % For multi-row cells in tables
\usepackage{amssymb} % For checkmarks, etc.
\usepackage[hyphens]{url} % For proper URL breaking
\usepackage{pifont} % For symbols like checkmarks \ding{51} \ding{55}
\usepackage{algorithm}
\usepackage{algpseudocode}
\title{Optimizing Domain-Adaptive Self-Supervised Learning for Clinical Voice-Based Disease Classification}
\name{Weixin Liu$^{\star}$, Bowen Qu$^{\star}$, Matthew Pontell$^{\dagger}$, Maria Powell$^{\dagger}$, Bradley Malin$^{\star, \dagger}$, Zhijun Yin$^{\star, \dagger}$}

\address{$^{\star}$ Vanderbilt University, Nashville, TN, USA\\
         $^{\dagger}$ Vanderbilt University Medical Center, Nashville, TN, USA \\
         Emails: \{weixin.liu, bowen.qu\}@vanderbilt.edu, \\
         \{matthew.e.pontell, maria.e.powell, b.malin, zhijun.yin\}@vumc.org}
\begin{document}
% Enable 9-point font size for ICASSP camera-ready version
\ninept
\maketitle
%
% ======================================================================
% --- Abstract ---
% ======================================================================
\begin{abstract}

The human voice is a promising non-invasive digital biomarker, yet deep learning for voice-based health analysis is hindered by data scarcity and %a challenge known as 
domain mismatch, where models pre-trained on general audio fail to capture the subtle pathological features characteristic of clinical voice data. To address these challenges, we investigate domain-adaptive self-supervised learning (SSL) with Masked Autoencoders (MAE) and demonstrate that standard configurations are suboptimal for health-related audio. Using the Bridge2AI-Voice dataset, a multi-institutional collection of pathological voices, we systematically examine three performance-critical factors: reconstruction loss (Mean Absolute Error vs. Mean Squared Error), normalization (patch-wise vs. global), and masking (random vs. content-aware). Our optimized design, which combines Mean Absolute Error (MA-Error) loss, patch-wise normalization, and content-aware masking, achieves a Macro F1 of $0.688 \pm 0.009$ (over 10 fine-tuning runs), outperforming a strong out-of-domain SSL baseline pre-trained on large-scale general audio, which has a Macro F1 of $0.663 \pm 0.011$. The results show that MA-Error loss improves robustness and content-aware masking boosts performance by emphasizing information-rich regions. These findings highlight the importance of component-level optimization in data-constrained medical applications that rely on audio data.

\end{abstract}

%
% ======================================================================
% --- Index Terms ---
% ======================================================================
%

% ======================================================================
% --- Section 1: Introduction ---
% ======================================================================
\section{Introduction}
\label{sec:intro}

The human voice, a rich and data-dense signal produced by the intricate coordination of respiratory, laryngeal, and articulatory systems, is emerging as a powerful non-invasive biomarker in digital health \cite{sara2023guess, bensoussan2024voice}. The acoustic properties of voice, such as pitch stability, harmonic structure, and airflow control, can reflect the underlying physiological and neurological state of an individual \cite{eyben2015geneva}. Consequently, voice analysis offers a low-cost, scalable, and remotely accessible modality for screening and monitoring a wide range of health conditions. Applications range from vocal fold disorders \cite{al2022effectiveness} and neurodegenerative diseases \cite{ali2024parkinson, yang2022deep}, to mood disorders \cite{tokuno2017pathophysiological, taguchi2018major} and respiratory conditions \cite{laguarta2020covid}.

Despite its promise, the application of deep learning to voice-based health analysis is hindered by two fundamental challenges. First, the acquisition of large-scale, high-quality labeled medical voice datasets is notoriously difficult due to patient privacy regulations, the high cost of clinical validation, and the requirement for expert annotation \cite{arasteh2024tackling}. This data scarcity limits the performance of traditional supervised learning models. Second, a common strategy to overcome data limitations is to employ transfer learning from models pre-trained on massive, open-source audio datasets like AudioSet \cite{gong2021ast, raghu2019transfusion, dubey2024convolutional}. However, this introduces a critical \textit{domain mismatch}. Informally, models pre-trained on general-purpose audio are optimized to distinguish between broad acoustic categories (e.g., speech vs. music), and their learned feature representations often lack the necessary resolution to capture the subtle, fine-grained acoustic variations, such as vocal tremor or hoarseness—that are indicative of specific pathologies \cite{gong2022ssast}. Furthermore, clinical voice data often presents significant complexity; for instance, the presence of \textit{comorbidity}, where a patient may have multiple simultaneous conditions, necessitates models capable of robust multi-label classification \cite{arasteh2024tackling}.

To address these limitations, specifically the critical bottleneck of expert annotation, we propose a domain-adaptive self-supervised learning (SSL) approach. The key advantage of SSL is its ability to learn robust representations directly from unlabeled data, thereby mitigating the dependency on large-scale, costly annotation efforts. While this paradigm does not resolve the underlying challenges of data privacy in collection, it effectively addresses the label scarcity problem. To implement this, we pre-train a model on unlabeled data drawn exclusively from the target domain of pathological voices. Specifically, we employ the Masked Autoencoder (MAE) paradigm \cite{he2022masked}, a powerful SSL technique, on the Audio Spectrogram Transformer (AST) \cite{gong2021ast} architecture. Our work is guided by a twofold central hypothesis: first, that standard MAE configurations, originally developed for general vision or audio tasks, are suboptimal for the unique, often non-stationary characteristics of pathological voice spectrograms, and second, that these configurations can be systematically optimized to unlock significant performance gains in this specialized domain.

This paper presents a systematic investigation to optimize the MAE framework for voice health analysis. To validate our methodology, we leverage the recently established Bridge2AI-Voice dataset \cite{Bensoussan2025_Bridge2AI}. Originating from the National Institutes of Health Bridge2AI initiative, this dataset represents a significant multi-institutional collection of voice recordings paired with clinical information, providing an ideal testbed for developing robust models. We deconstruct the MAE pre-training process into three components and conduct a comprehensive ablation study to evaluate their impact: 1) \emph{Reconstruction Loss Function}, 2) \emph{Input Normalization Strategy} (patch-wise vs. global), and 3) \emph{Masking Strategy} (content-aware vs. random). Our findings demonstrate that a carefully tuned domain-adaptive SSL model significantly outperforms a competitive generalist baseline, offering a methodological design reference for effectively adapting SSL to specialized and data-constrained medical audio domains.

% ======================================================================
% --- Section 2: Related Work ---
% ======================================================================
\section{Related Work}
\label{sec:related}

\subsection{Voice-based Health Analysis}
Traditional approaches to computational paralinguistics for health analysis have relied heavily on handcrafted feature sets, such as the eGeMAPS \cite{eyben2015geneva}. These features, which capture aspects like jitter, shimmer, and Harmonics-to-Noise Ratio (HNR), were fed into classical machine learning models like SVMs \cite{boyanov1997acoustic, martinez2000acoustic}. While interpretable, these features may not capture all relevant information \cite{arslan2024machine}. However, deep learning models operating on spectrograms have recently gained traction for their ability to learn hierarchical patterns automatically \cite{idrisoglu2023applied}.

\subsection{Self-Supervised Learning in Audio}
SSL has emerged as the primary method in audio processing to reduce label dependency. Models based on SSL can be broadly categorized into those operating on raw waveforms (e.g., Wav2Vec 2.0 \cite{baevski2020wav2vec}) and those operating on time-frequency representations such as spectrograms.
%\subsubsection{Audio Spectrogram Transformer (AST)}
 The Audio Spectrogram Transformer (AST) \cite{gong2021ast} adapted the Vision Transformer (ViT) architecture for audio classification. This model treats a log-mel spectrogram as an image, dividing it into a sequence of patches ($16 \times 16$ pixels). By applying Transformer encoders to this sequence, AST captures global dependencies across time and frequency, which has been shown to outperform prior CNN-based models on audio benchmarks. 
 
 The Self-Supervised AST (SSAST) \cite{gong2022ssast} extended the AST by incorporating self-supervised pre-training on the MAE framework. By pre-training on large general-purpose datasets (e.g., AudioSet, Librispeech) using a patch masking and reconstruction task, SSAST learns rich, general-purpose audio representations and thus can serve as a strong, leading baseline for audio SSL. However, the question remains whether these learned representations optimally transfer to the medical audio domain or if in-domain pre-training is superior.

% ======================================================================
% --- Section 3: Methodology for Domain-Adaptive Pre-training ---
% ======================================================================

\section{Methods}
\label{sec:method}

\subsection{Dataset and Problem Formulation}
\label{sec:dataset_baselines}
This study utilizes data from the Bridge2AI-Voice initiative \cite{Bensoussan2025_Bridge2AI}. The cohort consists of 442 unique participants recruited across five North American sites. After data cleaning and preprocessing, the final dataset consists of 16,738 distinct recordings.

The dataset encompasses a broad spectrum of health conditions, categorized into four primary types for this study: 1) Voice Disorders, 2) Neurological and Neurodegenerative Disorders, 3) Mood and Psychiatric Disorders, and 4) Respiratory Disorders. Table~\ref{tab:disease_distribution} provides a breakdown of the frequency of these conditions in the participants. As can be seen, the same participant can have multiple disorders. Consequently, we formulate the disease classification task as a multi-label classification problem, where each sample can possess multiple positive labels simultaneously.

\begin{table}[htbp]
\centering
\caption{Prevalence of disorders among the 442 study participants (percentages do not sum to 100\% due to comorbidity).}

\label{tab:disease_distribution}
\begin{tabular}{@{}lcc@{}}
\toprule
\textit{Disease Category} & \textit{Participant Count} & \textit{Percentage (\%)} \\
\midrule
Voice Disorders            & 230 & 52.0\% \\
Respiratory Disorders      & 226 & 51.1\% \\
Neurological Disorders     & 160 & 36.2\% \\
Mood/Psychiatric Disorders & 145 & 32.8\% \\
\bottomrule
\end{tabular}
\end{table}

The data contains pre-computed log-mel spectrograms for deep learning models, and 131 static features, including clinically-relevant acoustic features (e.g., jitter, shimmer) extracted via OpenSMILE \cite{eyben2010opensmile}, and prosodic features (e.g., pitch, formants) calculated using Praat and Parselmouth \cite{jadoul2018introducing}. The source audio was sampled at 16 kHz. Log-mel spectrograms were computed using a Short-Time Fourier Transform (STFT) with a 400-point FFT, corresponding to a 25 ms window size and a 10 ms hop length. The resulting representation was then mapped to 128 mel frequency bands ($n\_mels=128$) in a log-decibel scale. The primary spectrogram data used for training AST models consists of matrices with dimensions $128 \times N$ (mel bands $\times$ time frames). Given that raw audio waveforms (e.g., WAV files) were not available at the time of this study, we selected the AST architecture as it is designed explicitly on spectrogram patches, aligning perfectly with the provided data structure, unlike models like Wav2Vec 2.0 \cite{baevski2020wav2vec} which require raw waveform inputs.

\begin{figure}[t]
\centering
\includegraphics[width=0.45\textwidth]{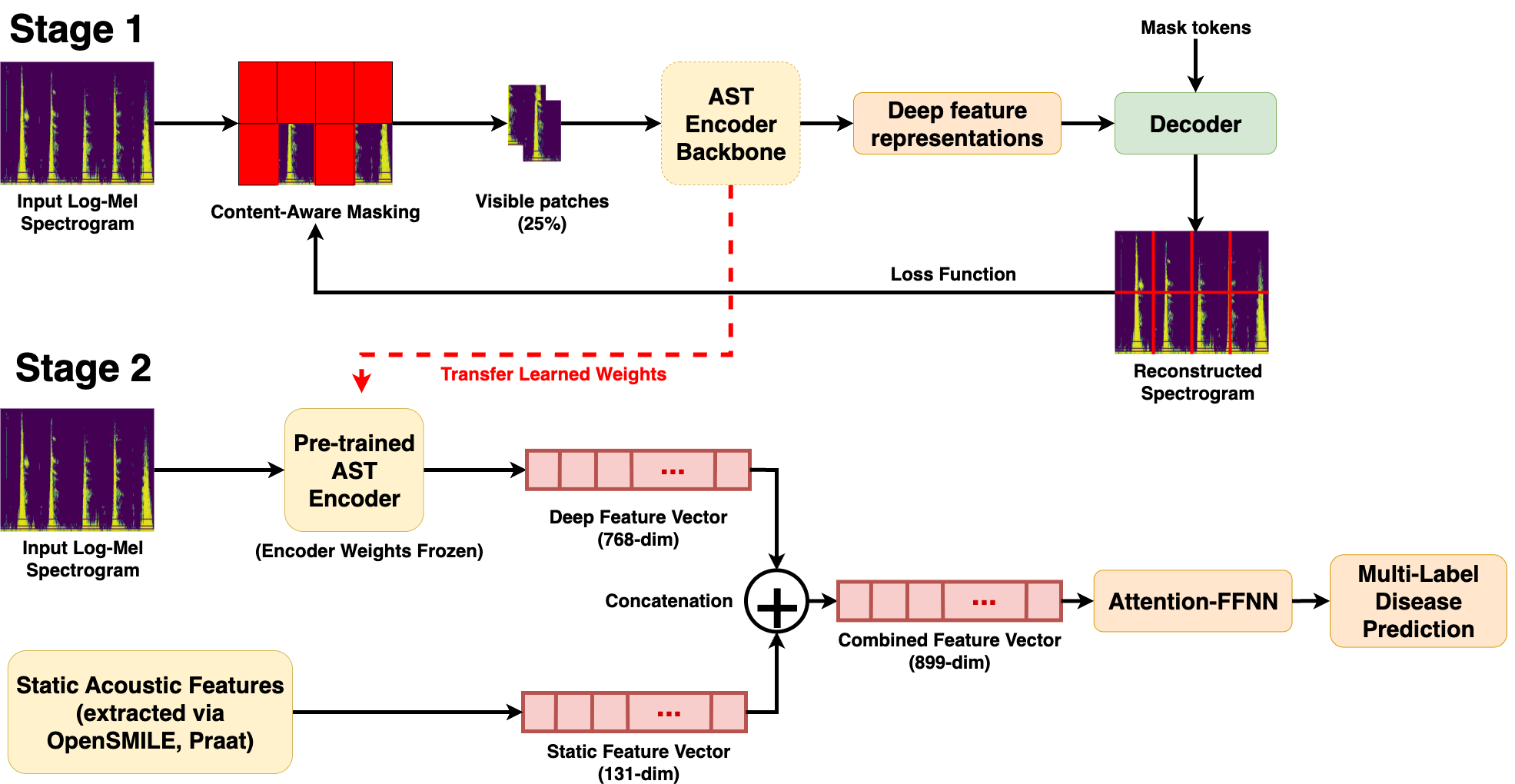} 
\caption{An overview of the two-stage framework. 
\textbf{Stage 1 (Domain-Adaptive SSL Pre-training):} An AST Encoder is pre-trained by reconstructing masked spectrogram patches, with its weights optimized via a reconstruction loss. 
\textbf{Stage 2 (Downstream Multi-Label Classification):} The learned encoder serves as a feature extractor, producing deep features that are fused with static acoustic features for the final prediction using an Attention-FFNN.
}
\label{fig:framework}
\end{figure}

Our two-stage methodology, illustrated in Fig.~\ref{fig:framework}, centers on optimizing the MAE pre-training process specifically for voice health data, followed by a downstream classification task.

% -----------------------------------------

\subsection{Stage 1: Domain-Adaptive SSL Pre-training}
In the pre-training stage, we adopt the asymmetric encoder-decoder architecture popularized by MAE \cite{he2022masked}. The core framework consists of several components. The first is \emph{Patch Embedding} where the input log-mel spectrogram $\mathbf{S} \in \mathbb{R}^{128 \times T}$ (frequency $\times$ time) is divided into a sequence of non-overlapping patches $\mathbf{p} \in \{\mathbf{p}_1, \mathbf{p}_2, \dots, \mathbf{p}_N\}$. Before feeding the patches into the model, each patch undergoes specific masking and normalization strategies.

\textit{Masking Strategy: Content-Aware vs. Random.}
We compare standard \emph{Random Masking} with \emph{Content-Aware Masking (CA)}, which focuses on high-saliency patches. For CA, we calculate saliency using \emph{patch variance} ($\sigma_{\mathbf{p}_i}^2$). Specifically, for a patch $\mathbf{p}_i$ with $K$ pixels (e.g., $K=256$ for a $16 \times 16$ patch), the variance is defined as:
$$ \sigma_{\mathbf{p}_i}^2 = \frac{1}{K} \sum_{j=1}^{K} (x_j - \mu_{\mathbf{p}_i})^2 $$
where $x_j$ is the value of a pixel within the patch and $\mu_{\mathbf{p}_i}$ is the mean pixel value of that patch. This metric can effectively measure signal complexity in spectrograms because pathological phenomena like jitter, shimmer, and harshness introduce high variability and complex harmonic structures, leading to higher patch variance. Compared to alternatives such as spectral entropy, patch variance strikes an optimal balance between identifying these critical pathological regions and maintaining high computational efficiency. Our implementation employs a \emph{hybrid strategy}: first, we define a high-saliency candidate pool by sorting all patches by variance. The pool size is dynamically set as $\max(0.7 \times N_{\text{masked}}, 0.5 \times N_{\text{total}})$ to ensure sufficient diversity. Second, 70\% of the mask quota is filled by sampling from this high-saliency pool, while the remaining 30\% is filled by random sampling from lower-saliency patches. This 70/30 split was selected after evaluating several candidate ratios (e.g., 50/50, 70/30, 80/20). We chose the ratio that maximized the downstream classifier's Macro F1-score on a held-out validation set. By forcing the model to infer complex regions from simpler context, we encourage a targeted understanding of acoustic interdependencies.

\textit{Input Normalization Strategy: Patch-wise Normalization.}
Audio recordings inherently have variations in volume and dynamic range. To counteract this, we test patch-wise normalization. Thus, each target patch $\mathbf{p}_i$ is normalized independently: $\mathbf{p}_{norm} = (\mathbf{p}_i - \mu_{\mathbf{p}_i}) / (\sigma_{\mathbf{p}_i} + \epsilon)$. The model's task becomes reconstructing this normalized patch. This decouples magnitude information from structural information.

Following the masking process, a high masking ratio of 0.75 is adopted. This ratio forces the model to learn a robust representation by reconstructing the signal from the remaining 25\% visible context, making the self-supervised task more challenging and effective. A standard AST \cite{gong2021ast} is applied as the \emph{Encoder} backbone, which processes only the visible patches. A lightweight Transformer \emph{Decoder} then processes the encoded patch tokens along with learnable `[MASK]' tokens representing the missing positions. Its objective is to reconstruct the original pixel values of the masked patches.

\textit{Reconstruction Loss Function: MA-Error vs. MSE.}
The loss function dictates how the model penalizes reconstruction errors. We evaluate two options rooted in different mathematical norms. The first is the Mean Absolute Error (MA-Error), often called L1 loss because it is based on the L1 norm of the error vector, defined as $\mathcal{L}_{\text{MA-Error}} = \frac{1}{|M|} \sum_{i \in M} \| \mathbf{p}_i - \hat{\mathbf{p}}_i \|_1$. Its use of absolute difference makes it inherently less sensitive to outliers. The second option is the Mean Squared Error (MSE), which is based on the squared L2 norm of the error vector, defined as $\mathcal{L}_{\text{MSE}} = \frac{1}{|M|} \sum_{i \in M} \| \mathbf{p}_i - \hat{\mathbf{p}}_i \|_2^2$. This loss, standard for many autoencoder frameworks, squares the error. This high penalty for large deviations forces the model to prioritize high-energy components (e.g., formants), which can be detrimental for pathology analysis where diagnostic information often resides in low-energy regions. We hypothesize that for heterogeneous pathological voice data, MA-Error encourages a more holistic representation of the patch structure.

\textit{Implementation Details.}
In the experiments, the decoder has an embedding dimension of 256, a depth of 4 layers, and 8 attention heads. The decoder is discarded after pre-training. All SSL models were pre-trained on the in-domain data for 120 epochs using the AdamW optimizer ($\beta_1=0.9$, $\beta_2=0.95$, weight decay=0.05). We applied a base learning rate of $1.5 \times 10^{-4}$ with a linear scaling rule for multi-GPU training, and utilized automatic mixed precision (AMP) to accelerate computation.

\subsection{Stage 2: Downstream Multi-Label Classification}
% After pre-training, the optimized AST encoder is used as a feature extractor for the downstream multi-label classification task. %, with its weights kept frozen. 
This stage utilizes a multi-modal feature fusion approach (see Fig.~\ref{fig:framework}). First, the pre-trained AST encoder processes an input log-mel spectrogram to produce a 768-dimensional deep feature vector. In parallel, a 131-dimensional static feature vector is prepared, comprising clinically-relevant acoustic features (e.g., jitter, shimmer) extracted via OpenSMILE \cite{eyben2010opensmile}, and prosodic features (e.g., pitch, formants) calculated using Praat and Parselmouth \cite{jadoul2018introducing}. These two vectors are then concatenated to form a combined feature vector. This combined vector is fed into an \emph{Attention-based Feed-Forward Network (Attention-FFNN)} for the final multi-label disease prediction. This classifier first employs a feature-level attention module to learn and assign an importance weight to each feature dimension. The re-weighted features are then processed by a multi-layer perceptron for the final classification.

% ==========================
% ======================================================================
% --- Section 4: Experimental Design ---
% ======================================================================
\section{Experimental Design}
\label{sec:exp_setup}
\subsection{Baseline Models}
We compare our domain-adaptive models against several baselines. The \emph{Supervised Baselines} include ResNet18 \cite{he2016deep} and EfficientNetB4, both pre-trained on ImageNet. We also include a \emph{General-Domain AST} model pre-trained on AudioSet \cite{gong2021ast}, and a strong \emph{Self-Supervised Baseline}, SSAST \cite{gong2022ssast}, pre-trained on AudioSet using SSL. To isolate the impact of the feature extractor and enable a direct comparison, we used a consistent downstream setup for all models. Each baseline model serves as a feature extractor, replacing the domain-adaptive pre-trained AST encoder, while all other components of the classification framework (described in Section 3.2) were held constant. The embedding vector produced by each baseline (e.g., 512-dim for ResNet18, 1,792-dim for EfficientNetB4, and 768-dim for AST-based models) is concatenated with the 131 static features and fed into the same Attention-FFNN classifier for the final prediction.

\subsection{Training and Evaluation}
We split the data at the participant level into training (80\%) and test (20\%) sets with no participant overlap. Hyperparameters (learning rates $\{1\mathrm{e}{-3}, 5\mathrm{e}{-4}, 1\mathrm{e}{-4}\}$; batch sizes $\{32, 64, 128\}$) were tuned via 5-fold cross-validation on the training set, and the best configuration was refit on the full training data. We used early stopping (patience = 10) and Focal Loss ($\gamma=2$) \cite{lin2017focal}. To quantify optimization-induced variance, we repeated downstream fine-tuning 10 times with different random seeds while keeping the train/test split fixed for our optimized model and SSAST. We report the mean Macro F1 in Table~\ref{tab:comprehensive_results}, while the corresponding standard deviations are reported in the main text for readability. Our primary metric is \emph{Macro F1}; we additionally report AUROC (Macro/Micro) and Micro F1 for completeness.

\begin{table*}[t]
\centering
\caption{Model performance comparison on the test set. \textbf{For our optimized model and SSAST, Macro F1 is averaged over 10 fine-tuning runs} with different random seeds (fixed train/test split); the corresponding standard deviations are reported in the main text.
Our optimized SSL configuration, using MA-Error loss, patch normalization (Norm), and content-aware (CA) masking, demonstrates superior performance over general-domain baselines and other ablation variants. 
For each metric, the best result is marked in \textbf{bold} and the second-best is \underline{underlined}}.
\label{tab:comprehensive_results}
\resizebox{\textwidth}{!}{%
\begin{tabular}{llccccccc}
\toprule
\textbf{Model Group} & \textbf{Model Configuration} & \textbf{Macro F1} & \textbf{Macro AUC} & \textbf{Macro Accuracy} & \textbf{Micro F1} & \textbf{Macro Precision} & \textbf{Macro Recall} & \textbf{Micro AUC} \\
\midrule
% --- Optimized Model ---
\multirow{1}{*}{\textbf{Our Optimized SSL}} & \textbf{SSL-AST (MA-Error + Norm + CA)} & \textbf{0.688} & \underline{0.813} & \textbf{0.767} & \textbf{0.726} & \underline{0.839} & \textbf{0.633} & \textbf{0.852} \\
\midrule
% --- Key Baselines ---
\multirow{5}{*}{Baselines} & SSAST (Pre-trained on AudioSet) & \underline{0.663} & 0.791 & \underline{0.758} & 0.711 & 0.791 & 0.613 & 0.825 \\
& AST (Pre-trained on AudioSet) & 0.624 & 0.774 & 0.728 & 0.667 & 0.743 & 0.562 & 0.815 \\
& Static features only (131-d) & 0.619 & 0.770 & 0.732 & 0.661 & 0.749 & 0.553 & 0.824 \\
& ResNet18 (Pre-trained on ImageNet) & 0.610 & \textbf{0.814} & 0.742 & 0.676 & 0.817 & 0.550 & \underline{0.845} \\
& EfficientNetB4 (Pre-trained on ImageNet) & 0.563 & 0.800 & 0.714 & 0.622 & 0.819 & 0.480 & 0.827 \\
\midrule
% --- Ablation Study Models (Ranked) ---
\multirow{7}{*}{SSL-AST Ablations} & MA-Error + CA & 0.655 & 0.785 & \underline{0.758} & \underline{0.713} & 0.816 & \underline{0.614} & 0.829 \\
& MSE + Norm + CA & 0.641 & 0.768 & 0.736 & 0.689 & 0.746 & 0.599 & 0.815 \\
& MSE + CA & 0.622 & 0.786 & 0.725 & 0.667 & 0.751 & 0.567 & 0.831 \\
& MA-Error + Norm & 0.611 & 0.786 & 0.736 & 0.674 & \textbf{0.842} & 0.560 & 0.827 \\
& MSE + Norm & 0.609 & 0.781 & 0.730 & 0.676 & 0.760 & 0.567 & 0.818 \\
& MA-Error (Base) & 0.608 & 0.791 & 0.736 & 0.674 & 0.764 & 0.555 & 0.829 \\
& MSE (Base) & 0.592 & 0.777 & 0.733 & 0.660 & 0.809 & 0.524 & 0.822 \\
\bottomrule
\end{tabular}%
}
\end{table*}

% ======================================================================
% --- Section 5: Results and Discussion ---
% ======================================================================
\section{Results}
\label{sec:results}

\subsection{Overall Performance Comparison}
Table~\ref{tab:comprehensive_results} reports the performance of all models on the held-out test set. Overall, our optimized SSL model achieves the best performance across all metrics, except Macro AUC. While ResNet18 has the best Macro AUC, the best baseline model is SSAST, which has the second-best performance in Macro F1 and Macro Accuracy. To account for optimization-induced variance, we repeated the downstream fine-tuning task 10 times with different random seeds (fixed train/test split) for our optimized model and SSAST, a strong baseline. Our optimized model, \emph{SSL-AST (MA-Error+Norm+CA)}, achieves the best Macro F1 of $0.688 \pm 0.009$, outperforming the \underline{second-best} model in Macro F1, SSAST pre-trained on large-scale general audio, which attains $0.663 \pm 0.011$. 
This absolute gain of $0.025$ in Macro F1 supports our hypothesis that tailoring MAE-style SSL to the clinical voice domain yields more diagnostically useful representations than relying on larger out-of-domain pre-training.

\subsection{Analysis of Component Contributions}
\label{sec:discussion_component_analysis}
Here, we present the contribution of each optimization component.

\textit{Impact of Loss Function (MA-Error vs. MSE).} MA-Error loss consistently outperforms MSE across all configurations. For instance, our best model using MA-Error loss (MA-Error+Norm+CA) achieved a Macro F1 of $0.688 \pm 0.009$, surpassing the best-performing MSE counterpart (MSE+Norm+CA) at 0.641, a 7.3\% relative improvement. While this can be explained in part by MA-Error’s robustness against extreme values, a more important factor is how it balances attention between dominant and subtle cues. MSE, by squaring the error term, heavily penalizes large deviations and biases learning toward high-energy components such as stable formants. However, pathological voice quality is often defined by low-energy, irregular features (e.g., breathiness, turbulence, or aperiodic bursts) that carry crucial diagnostic information but contribute little energy to the spectrogram. By weighting errors more evenly, MA-Error encourages the model to capture both dominant formants and subtle pathological cues, which produces a more holistic acoustic representation.

\textit{Impact of Masking Strategy (Content-Aware vs. Random).}
Content-aware (CA) masking yields the largest single performance gain. For instance, holding other factors constant, adding CA masking to the base model with MA-Error loss improved its Macro F1 from 0.608 to 0.655 (the MA-Error+CA model), marking a 7.7\% relative increase. This supports the idea that the difficulty of the reconstruction task matters. In spectrograms, high-variance patches typically capture dynamic acoustic events—voiced/unvoiced transitions or regions of harmonic instability. Masking these regions forces the model to reconstruct challenging events from context, promoting learning of deeper interdependencies in vocal production. As a result, the model develops richer, diagnostically relevant representations than with random masking, which often removes only stable, easily interpolated regions.

\textit{Synergy and interpreting Macro AUC vs. Macro F1.}
Patch normalization consistently improves performance. For example, adding normalization to our MA-Error+CA model increases Macro F1 from 0.655 to 0.688, demonstrating its effectiveness. Overall, the best performance is achieved by combining all three components (MA-Error loss, patch normalization, and CA masking), highlighting their synergy. We also report Macro AUC as a complementary, threshold-independent metric (Table~\ref{tab:comprehensive_results}). While Macro AUC reflects ranking ability across decision thresholds, Macro F1 evaluates performance at a fixed operating threshold (0.5 in our experiments). Consistent with the Macro F1 results, our optimized model also achieves competitive Macro AUC compared with SSAST (the \underline{second-best} model in Macro F1), indicating strong ranking ability as well as strong thresholded performance.

% \textit{Clinical Implications and Limitations}
% The finding that a smaller in-domain dataset can yield superior results to a massive general-dataset, given proper SSL optimization, has significant implications for computational health research. This is particularly relevant for applications involving rare diseases where data collection is inherently limited, suggesting that investment in sophisticated, domain-specific pre-training strategies can be more fruitful than relying solely on larger, out-of-domain resources.

% However, we acknowledge limitations. While the Bridge2AI cohort is comprehensive, a sample size of 442 unique participants for deep learning still necessitates careful validation to ensure generalizability across different populations and recording conditions. Furthermore, our focus on Macro F1 over Macro AUC, while clinically justified for a screening tool (where balancing precision and recall at a specific operating point is crucial), highlights the need to consider multiple metrics and application contexts when evaluating diagnostic models.

% ======================================================================
% --- Section 6: Conclusion ---
% ======================================================================
\section{Discussions and Conclusion}
\label{sec:conclusion}

We presented a systematic optimization of domain-adaptive SSL for voice-based disease classification under domain mismatch. Through an ablation study, we evaluated how reconstruction loss, input normalization, and masking strategy affect MAE-style spectrogram pre-training. Our main finding is that an optimized configuration---\emph{MA-Error} loss with patch-wise normalization and content-aware masking---pre-trained on in-domain pathological voices, outperforms strong SSL baselines trained on much larger general-purpose corpora, indicating that objective-level tailoring can be more effective than scaling out-of-domain data. Despite Bridge2AI being multi-institutional, the cohort size (442 participants) warrants caution when generalizing across populations and recording conditions. Future work will study end-to-end models, interpretability \cite{gawlikowski2023survey}, and cross-dataset generalization \cite{arasteh2024tackling}. Overall, objective-level tailoring is a practical alternative to scaling out-of-domain data.

\textbf{Acknowledgment.} Maria Powell, Bradley Malin, and Zhijun Yin are supported by NIH/NHGRI grant U54HG012510. The authors declare no conflicts of interest.

\textbf{Compliance with Ethical Standards.} This work is a secondary analysis of the de-identified Bridge2AI-Voice dataset accessed under the Bridge2AI Voice Registered Access License and Data Use Agreement; the authors had no access to direct identifiers, did not attempt re-identification, and performed no interaction or intervention with human participants.

% ===================================
% ======================================================================
% --- References ---
% ======================================================================

% References should be produced using the bibtex program from suitable
% BiBTeX files (here: strings, refs, manuals). The IEEEbib.bst bibliography
% style file from IEEE produces unsorted bibliography list.
% -------------------------------------------------------------------------
\clearpage
\bibliographystyle{IEEEbib}
\bibliography{references} % Assumes your .bib file is named references.bib

\end{document}